\documentclass[conference]{IEEEtran}
\usepackage{graphicx}
\usepackage{subfigure}
\usepackage{float}
\usepackage{algorithm}
\usepackage{algorithmic}
\usepackage{flushend}
\usepackage{mathrsfs}
\usepackage{amsfonts}
\usepackage{array}
\usepackage{amssymb,amsmath}
\usepackage{longtable}
\usepackage{rotating}
\usepackage{multirow}
\usepackage{epstopdf}
\usepackage{flushend}
\usepackage{fancyhdr}
\usepackage{cite}
\usepackage{float}
\usepackage[marginal]{footmisc}
\usepackage{stfloats}
\usepackage[colorlinks,
            linkcolor=black,
            anchorcolor=black,
            citecolor=black]{hyperref}
\newcommand{\trace}{{\rm tr}}
\renewcommand{\baselinestretch}{1}

\def\BibTeX{{\rm B\kern-.05em{\sc i\kern-.025em b}\kern-.08em
    T\kern-.1667em\lower.7ex\hbox{E}\kern-.125emX}}

\renewcommand{\baselinestretch}{1}
\newtheorem{theorem}{Theorem}

\newtheorem{lemma}{Lemma}

\begin{document}

\title{Reconfigurable Intelligent Surface for MISO Systems with Proportional Rate Constraints}
\author{\IEEEauthorblockN{Yulan Gao\textsuperscript{1,2}, Chao Yong\textsuperscript{1}, Zehui Xiong\textsuperscript{2}, Dusit Niyato\textsuperscript{2},  Yue Xiao\textsuperscript{1}, Jun Zhao\textsuperscript{2}}

\IEEEauthorblockA{\textsuperscript{1}{The National Key Laboratory of Science
and Technology on Communications} \\
{University of Electronic
Science and Technology of China, Chengdu 611731, China } }

\IEEEauthorblockA{\textsuperscript{2}{The School of Computer Science and Engineering,}
{Nanyang Technological University, Singapore  639798}}

\IEEEauthorblockA{yulangaomath@163.com, chaoyongmath@163.com, xiaoyue@uestc.edu.cn, \{zxiong002, DNiyato, JunZhao\}@ntu.edu.sg}
}

\maketitle

\thispagestyle{fancy}
\pagestyle{fancy}
\lhead{This paper appears in the Proceedings of IEEE International Conference on Communications (ICC) 2020.\\ Please feel free to contact us for questions or remarks.}
\cfoot{\thepage}
\renewcommand{\headrulewidth}{0.4pt}
\renewcommand{\footrulewidth}{0pt}
\begin{abstract}
This paper investigates the spectral efficiency (SE) in reconfigurable intelligent surface (RIS)-aided multiuser multiple-input single-output (MISO) systems, where RIS can reconfigure the propagation environment via a large number of controllable and intelligent phase shifters.
In order to explore the SE performance with user proportional fairness for such a system, an optimization problem is formulated to maximize the SE by jointly considering the power allocation at the base station (BS) and phase shift at the RIS, under nonlinear proportional rate fairness constraints.
To solve the non-convex optimization problem, an effective solution is developed, which capitalizes on an iterative algorithm with closed-form expressions, i.e., alternatively optimizing the transmit power at the BS and the reflecting phase shift at the RIS.
Numerical simulations are provided to validate the theoretical analysis and assess the performance of the proposed alternative algorithm.
\end{abstract}
\IEEEpeerreviewmaketitle

\section{Introduction}
Large antenna systems as a key technique for the fifth-generation (5G) and future mobile communications, which promise to scale up the performance of  conventional communication systems by utilizing the spatial degrees of freedom.
However, owing to the existence of buildings, trees, cars, and even humans, the obstacles still occur in  large-scale antenna systems \cite{Han2018Large}.
Recently, a promising way to address this problem is the reconfigurable intelligent surfaces (RIS) which can proactively reconfigure the wireless propagation environment \cite{Renzo2019Smart}.
Specifically, RIS is installed with a large number of low-cost reflecting elements, which are controllable and intelligent to induce an amplitude and phase change to the incident signal independently.
From an operational standpoint, RIS can be integrated into existing fundamental wireless infrastructures and buildings, regarded as a complement to existing wireless communication networks.

Available works on RIS-aided communication systems mainly focused on the system spectral efficiency (SE) and receive signal power performance.
Among the early contribution in this area, Ref. \cite{Wu2018Intelligent} investigated the received signal power maximization problem for RIS enhanced multiple-input single-output (MISO) systems by jointly optimizing the transmit beamforming at the base station (BS) and reflect the beam by the phase shifter at the RIS.
In the spirit of these works, a vast corpus of literature has  focused on developing  active-passive beamforming design techniques for SE maximization, transmit power minimization, and energy efficiency optimization, etc., subject to maximum transmit power and minimum quality of service (QoS) constraints.
For instance, Ref. \cite{Wu2019Intelligentjournal} proposed alternating optimization techniques for cost effect beamforming design in downlink MISO systems, Ref. \cite{Guo2019Weighted} investigated the unilateral SE maximization problem, while global energy efficiency   was studied in \cite{Huang2019Reconfigurable}.
These resource management problems mentioned for performance maximization only consider the maximum transmit power or minimum QoS constraints, which are designed from the perspective of the entire system without considering the fairness requirements of different users.
Consequently, optimization problem formulation that focuses exclusively on the entire system perspective is not aligned with different user fairness requirements, which as mentioned in \cite{Xiong2016Energy}, is crucial for next generation wireless networks.

In this paper, we investigate the SE performance of RIS-aided MISO systems, where a BS transmits signals to multiple users with the help of RIS.
Different from the existing works, in this paper, SE performance is maximized with the proportional rate fairness among multiple users by considering a set of nonlinear rate ratio constraints.
Since the downlink fairness is critical for supporting various multimedia applications in future mobile communication systems \cite{Yang20196G},
and the proportional rate constraints can guarantee the instantaneous fairness of multiple users.
Therefore, the main contributions of this paper are summarized as follows.
We consider a balanced tradeoff between SE and user fairness in RIS-aided systems.
Specifically, our goal is to maximize the SE via joint transmit power allocation at the BS and phase shift at the RIS while guaranteeing the predetermined fairness criterion.
An iterative algorithm with closed-form expressions is proposed to alternatively optimize the transmit power allocation and the phase shift.
Simulations show that the optimal solution based on the alternative method achieves significant performance gain compared with various baselines, such as the random phase shift method and the non-RIS zero-forcing (ZF) transmission method.

{\sl Notation:}
Matrices and vectors are denoted by bold letters.
${\mathbb E}\{\cdot\}$ denotes expectation.
$\triangleq $ denotes equality by definition.
We use $\|\cdot\|_F$ to denote the Frobenius norm of a matrix.
$\otimes$ stands for the Kronecker product of matrices.
We use ${\bf A}\succeq {\bf B}$ to indicate that ${\bf A}-{\bf B}$ is a positive semi-defined matrix.
$\text{diag}(\bf a)$ denotes the diagonal matrix with diagonal entries consisting the elements of vector ${\bf a}.$
$\trace(\cdot)$ denotes the trace of a matrix.
$\text{vec}(\bf A)$ is a vector stacking all the columns of matrix ${\bf A}.$
Real part, modulus, conjugate, and the angle of a complex number $a$ are denoted by
$\text{Re}(a), |a|, (a)^{\dag},$ and $\arg(a)$, respectively.
We use $(\cdot)^T, (\cdot)^{H}, (\cdot)^{-1},$ and $ (\cdot)^{\ddag}$ to denote transpose, Hermitian, inverse, and matrix pseudo-inverse, respectively.
Notation ${\bf x}\sim{\cal C}{\cal N}(0,\sigma^2)$ means that random variable ${\bf x}$ is complex circularly symmetric Gaussian with zero mean and variance $\sigma^2.$
${\mathbb R}$ and ${\mathbb C}$ denote the complex and real number sets, respectively.

\section{System Model and Problem Formulation}
\subsection{System Model}

\begin{figure}[!t]
\centering
\begin{minipage}[t]{0.7\linewidth}
\centering
\includegraphics[width=1\linewidth]{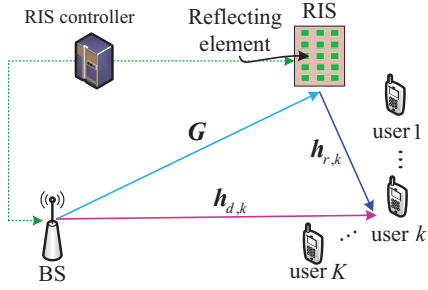}
\end{minipage}
\caption{An RIS-aided multiuser MISO system.}
\label{fig:0}
\end{figure}

As shown in Fig.\ref{fig:0}, we consider the RIS-aided multiuser MISO communication system consisting of a BS equipped with $M\text{-antennas}$, one RIS installed with $N$ reflecting elements, and a set ${\cal K}\triangleq \{1, 2, \ldots, k, \ldots, K\}$ of $K$ users, and we assume $M\geq K.$
Let ${\bf G}\in {\mathbb C}^{N\times M}$ denote the complex channel matrix between the BS and RIS,
${\bf h}_{r,k}\in {\mathbb C}^{N\times 1}$ denote the complex channel coefficient vector between the RIS and user $k, \forall k\in {\cal K},$  and the direct link from the BS to the $k\text{th}$ user is ${\bf h}_{d,k}\in {\mathbb C}^{M\times 1}.$

Then, the received signal at user $k$ is given by
\begin{equation}\label{eq:1}
\begin{aligned}
y_k=\left({\bf h}_{d,k}^{H}+{\bf h}_{r, k}^{H}{\pmb\Phi}^{H}{\bf G}\right){\bf x}+u_k, \forall k\in {\cal K},
\end{aligned}
\end{equation}
where ${\pmb\Phi}\triangleq \sqrt{\eta}\text{diag}[\phi_1, \phi_2, \ldots, \phi_N]$ is a diagonal matrix introduced by the effective phase shifts of all RIS reflecting elements, and $\eta$  indicates the reflection efficiency,
while $u_k\sim{\cal C}{\cal N}(0, \sigma^2)$ models the terminal additive white Gaussian noise (AWGN)
at receiver $k$.
Moreover, ${\bf x}=\sum_{k=1}^K \sqrt{p_k}{\bf w}_ks_k$ is the transmitted signal at the BS, $s_k$ is the original signal intended for the $k\text{th}$ user satisfying ${\mathbb E}\{|s_k|^2\}=1,$ $p_k$ represents the transmit power at the BS, and ${\bf w}_k\in {\mathbb C}^{M\times 1}$ is the precoding vector.
The transmit power at the BS should satisfy the constraint, as
\begin{equation}\label{s:1}
{\mathbb E}\{|{\bf x}|^2\}=\trace({\bf P}{\bf W}^H{\bf W})\leq P_{\max},
\end{equation}
where $P_{\max}$ is the maximum allowable transmit power,
${\bf W}\triangleq [{\bf w}_1, {\bf w}_2, \ldots, {\bf w}_K]\in {\mathbb C}^{M\times K}, $ and
${\bf P}\triangleq \text{diag}[p_1, p_2, \ldots, p_K]\in {\mathbb C}^{K\times K}.$

Based on (\ref{eq:1}), the receive signal-to-interference-and-noise ratio (SINR)
at user $k$ is denoted as
\begin{equation}\label{s:2}
\gamma_k\triangleq\frac{p_k\left|\left( {\bf h}_{d,k}^H+{\bf h}_{r,k}^H{\pmb\Phi}{\bf G}      \right){\bf w}_k\right|^2}
{\sum_{i=1,i\neq k}^K p_i\left|\left({\bf h}_{d,k}^H+{\bf h}_{r,k}^H {\pmb\Phi}{\bf G}  \right){\bf w}_i\right|^2+\sigma^2}.
\end{equation}
Then, the SE of the $k\text{th}$ user is obtained via Shannon formula as
\begin{equation}\label{s:3}
R_k=\log_2\left(  1+\gamma_k \right), \forall k\in {\cal K}.
\end{equation}

\subsection{Problem Formulation}
Our goal is to jointly optimize the transmit power allocation at the BS and the effective phase shift for the RIS elements to maximize SE for the RIS-aided MISO system, while satisfying the BS transmit power constraint $P_{\max}$.
Moreover, we introduce the idea of proportional fairness into the system by adding a set of nonlinear rate ratio constraints.
Besides, we consider the ideal reflection coefficient for the RIS elements, appearing in the diagonal of ${\pmb\Phi}=\sqrt{\eta}\text{diag}[\phi_1, \phi_2, \ldots, \phi_N],$ the peak-power constrained reflection coefficient is $|\phi_n|\leq 1, \forall n=1, 2, \ldots, N.$
Mathematically, SE optimization problem is formulated as
\begin{subequations}\label{add:1}
\begin{align}
&\max_{{\bf P}, {\pmb\Phi}}~  \sum_{k=1}^K \log_2\left(1+\gamma_k\right)\\\
\text{s.t.}&~ \trace({\bf P}{\bf W}^H{\bf W})\leq P_{\max},\\
&~ |\phi_n|\leq 1, \forall n=1, 2, \ldots, N,\\
&~ R_1: R_2:\cdots: R_K=\xi_1:\xi_2:\cdots:\xi_K,
\end{align}
\end{subequations}
where (\ref{add:1}d) is the proportional user rate constraint,
in which $\{\xi_1, \xi_2, \ldots, \xi_K\}$ is the set of predetermined proportional coefficients that are used to ensure fairness among the users \cite{Xue2012Radio}.
In other words, the data rate service of the users is performed with a quantized priority.
Thus, unlike other global SE (i.e., the sum-rate) \cite{Wu2018Intelligent} and Sum-SE (i.e., the weighted sum-rate)\cite{Guo2019Weighted} maximization problems, the transmit beamforming matrix and phase shift are tackled in a joint manner from the entire system perspective.
Furthermore, the proportional rate constraints indicates that the data rate among the $K$ users should follow a predetermined proportion and, in doing so, obviously affects the strategies of BS and RIS.

Note that problem (\ref{add:1}) is neither convex nor quasi-convex due to the non-convex objective function w.r.t. ${\bf P}$ and ${\pmb\Phi}$, and the nonlinear proportional rate constraint (\ref{add:1}d).
In general, there is no  effective and standard method to solve such kind of problems optimally.
Therefore, we aim to propose a distributed algorithm for solving (\ref{add:1}) by applying alternating optimization techniques.
To make the targeted problem more tractable, we assume that all involved channels are perfectly known at BS that employ ZF transmission, which is known to be optimal in the high SINR regime \cite{Wagner2013Large,Huang2018Asymptotically}.
To this end, it is assumed that the BS perfectly knows all the communication channels ${\bf h}_{d,k}, {\bf h}_{r,k}, $ and ${\bf G}$, which can be acquired by the methods described in \cite{Alkhateeb2018Deep}.

Denote the combined channel for user $k$ by ${\bf h}_k={\bf h}_{d,k}+{\bf G}^H {\pmb\Phi}^{H} {\bf h}_{r,k}, \forall k\in {\cal K}.$
Notably, the precoding matrix ${\bf W}$ should be designed based on the ZF criterion to cancel the interference from other users.
Specifically, after the transmit ZF processing, the received signal at user $k$ is
$y_k=\sqrt{p_k}s_k~| u_k=0, \forall k\in {\cal K}$, which means if the receiver noise $u_k=0$, the received signal by the transmit ZF processing with ${\bf W}$ should be exactly equal to $\sqrt{p_k}s_k.$
Define ${\bf H}_1\triangleq [{\bf h}_{d,1},  \ldots, {\bf h}_{d, K}]\in {\mathbb C}^{M\times K}$
and ${\bf H}_2=[{\bf h}_{r,1},  \ldots, {\bf h}_{r, K}]\in {\mathbb C}^{N\times K}.$
Then, the perfect interference suppression is achieved by setting the ZF precoding matrix to
${\bf W}=\left( {\bf H}_1^H+{\bf H}_2^H {\pmb\Phi}{\bf G} \right)^{\ddag}.$
Then,  SE maximization problem can be formulated as
\begin{subequations}\label{s:4}
\begin{align}
&\max_{{\bf P}, {\pmb\Phi}} ~\sum_{k=1}^K \log_2\left(1+p_k\sigma^{-2}\right)\\
\text{s.t.} & ~\trace\left(\left({\bf H}_1^H+{\bf H}_2^H {\pmb\Phi}{\bf G}\right)^{\ddag}\right.\nonumber\\
&~~~~~~~\left.{\bf P}\left({\bf H}_1^H+{\bf H}_2^{H}{\pmb\Phi}{\bf G} \right)^{\ddag H}\right)\leq P_{\max},\\
& ~~~~(\ref{add:1}\text{c}) \text{~and~}  (\ref{add:1}\text{d}).
\end{align}
\end{subequations}

\section{SE Maximization }

The difficulty of problem (\ref{s:4}) are two-fold: 1) the set defined by constraint (\ref{s:4}b) is non-convex; 2) the nonlinear proportional rate fairness constraints (\ref{add:1}d) are introduced.
In the following, we first consider two sub-problems of (\ref{s:4}), namely phase shift optimization with fixed transmit power allocation at the BS and transmit power allocation optimization with fixed  ${\pmb\Phi}$.
Iterating this process improves the system performance at each iteration step, and must eventually converge in the optimum value of the objective.
In the rest of this section, the optimization with respect to $\pmb\Phi$ for fixed ${\bf P},$
and with respect to ${\bf P}$ for fixed $\pmb\Phi$ will be treated separately.

\subsection{Optimization  ${\pmb\Phi}$ with Fixed $\bf P$}
In order to solve the nonlinear proportional user rate constraints in (\ref{add:1}d), we introduce an intermediate variable $\varphi,$
which is defined as
\begin{equation}
\varphi=\frac{R_1}{\xi_1}=\frac{R_2}{\xi_2}=\cdots=\frac{R_K}{\xi_K}.
\end{equation}
Hence,  SE is reformed as $\varphi\sum_{k=1}^K \xi_k,$ and the objective of (\ref{s:4}) is converted into seeking the maximum $\varphi.$
On the other hand, power requirement of user $k$ can be expressed based on (\ref{s:4}a) as
\begin{equation}\label{eqq:1}
p_k(\varphi)=\sigma^2\left(2^{\varphi\xi_k}-1\right), \forall k\in {\cal K}.
\end{equation}
Hence, with the following definition
\begin{equation}\label{s:7}
{\bf P}(\varphi)=\text{diag}[p_1(\varphi), p_2(\varphi), \ldots, p_K(\varphi)],
\end{equation}
the optimization problem (\ref{s:4}) can be rewritten as follows:
\begin{subequations}\label{s:8}
\begin{align}
&\max_{\varphi, {\pmb\Phi}}~ \varphi\\
\text{s.t.~}&\trace\left(({\bf H}_1^H+{\bf H}_2^{H}{\pmb\Phi}{\bf G})^{\ddag}{\bf P}(\varphi)\right.\nonumber\\
&~~~~~\left.({\bf H}_1^H+{\bf H}_2^H{\pmb\Phi}{\bf G})^{\ddag H} \right)\leq P_{\max},\\
&\hspace{6mm}\text{and ~} (\ref{add:1}\text{c}).
\end{align}
\end{subequations}

For a fixed transmit power allocation matrix ${\bf P},$ i.e., the intermediate variable $\varphi$ is fixed, the objective can be regarded as a constant value objective function $\varphi$ under constraints (\ref{add:1}c) and (\ref{s:8}b).
In this context, (\ref{s:8}) can be shown to equivalent to be the following problem:
\begin{equation}\label{s:9}
\begin{aligned}
&\max_{\pmb\Phi}~ \varphi\\
~~\text{s.t.}&~ (\ref{add:1}\text{c})~ \text{and~} (\ref{s:8}\text{b}).
\end{aligned}
\end{equation}
In our setting, note that the challenge for solving the optimization problem (\ref{s:9}) is the non-differentiability of its objective function.
To proceed further, we observe that (\ref{s:9}) is feasible if and only if the solution of the following optimization problem
\begin{equation}\label{s:10}
\begin{aligned}
&\min_{\pmb\Phi} \trace\left(({\bf H}_1^H+{\bf H}_2^{H}{\pmb\Phi}{\bf G})^{\ddag}{\bf P}(\varphi)
({\bf H}_1^H+{\bf H}_2^H{\pmb\Phi}{\bf G})^{\ddag H} \right)\\
\text{s.t.~}& |\phi_n|\leq 1, \forall n=1, 2, \ldots, N,
\end{aligned}
\end{equation}
is such that the objective can be made lower than $P_{\max}.$
\begin{theorem}
The optimization problem (\ref{s:10}) with respect to ${\pmb\Phi}$ for given $\varphi$ (i.e., ${\bf P}$) can be rewritten as
\setcounter{equation}{12}
\begin{subequations}\label{s:11}
\begin{align}
&\min_{\pmb\Phi}~ \text{vec}({\pmb\Phi}^{-1})^H{\bf A}\text{vec}({\pmb\Phi}^{-1})\nonumber\\
&~~~~~~~+2\text{Re}\left\{\text{vec}({\bf G}\overline{\bf H}_1^{\ddag}\overline{\bf H}_2)^H{\bf A}\text{vec}(\pmb\Phi^{-1})\right\}\\
\text{s.t.}&~ |\phi_n|\leq 1, \forall n=1, 2, \ldots, N,
\end{align}
\end{subequations}
where ${\bf A}=(\overline{\bf H}_2^{\ddag H}\otimes{\bf G}^{\ddag})^H(\overline{\bf H}_2^{\ddag H}\otimes {\bf G}^{\ddag}),$  $\overline{\bf H}_1={\bf Q}{\bf H}_1^H,$ and $ \overline{\bf H}_2={\bf Q}{\bf H}_2^{H}$ with ${\bf Q}{\bf Q}^H={\bf P}.$
\end{theorem}

{\textit{ Proof:}} The proof of Theorem 1 can be found in the Appendix A of this paper.

However, $|\phi_n|^2\leq 1, $ is not a complex analytic function. Thus, we rewrite the constraint (\ref{s:11}b) as
\begin{equation}\label{s:13}
\begin{aligned}
\text{vec}(\pmb\Phi^{-1})^H&e_{n+N(n-1)}e_{n+N(n-1)}^H\text{vec}(\pmb\Phi^{-1})\leq 1,
\end{aligned}
\end{equation}
where $e_{n+N(n-1)}\in {\mathbb R}^{N^2\times 1}$ is an elementary vector with a one at the $(n+N(n-1))\text{th}$ position. Then, the optimization problem (\ref{s:11}) is represented as
\begin{subequations}\label{s:14}
\begin{align}
&\min_{\pmb\Phi}~ \text{vec}({\pmb\Phi}^{-1})^H{\bf A}\text{vec}({\pmb\Phi}^{-1})\nonumber\\
&~~~~~~~~+2\text{Re}\left\{\text{vec}({\bf G}\overline{\bf H}_1^{\ddag}\overline{\bf H}_2)^H{\bf A}\text{vec}(\pmb\Phi^{-1})\right\}\\
\text{s.t.}&~ \text{vec}(\pmb\Phi^{-1})^He_{n+N(n-1)}e_{n+N(n-1)}^H\text{vec}(\pmb\Phi^{-1})\leq 1.
\end{align}
\end{subequations}

Since the zero duality gap holds for the above problem (\ref{s:14}), the optimal phase shift matrix $\pmb\Phi$ can be obtained through the Lagrange dual domain.
Let ${\pmb\lambda}=[\lambda_1, \lambda_2, \ldots, \lambda_N]^T$ denote the Lagrange multipliers associated with phase shift constraints (\ref{s:13}).
The dual problem of the optimization problem (\ref{s:14}) is given by
\begin{equation}\label{s:15}
\begin{aligned}
&\max_{\pmb\lambda}\min_{{\bf y}}~{\cal L}({\pmb\lambda}, {\bf y})\\
\text{s.t.}&~\pmb\lambda\succeq 0,
\end{aligned}
\end{equation}
where ${\bf y}=\text{vec}(\pmb\Phi^{-1}),$   and the Lagrangian function in (\ref{s:15}) is expressed as
\begin{equation}\label{eeq:1}
\begin{aligned}
{\cal L}({\pmb\lambda}, {\bf y})=&\left\{{\bf y}^H{\bf A}{\bf y}+2\text{Re}\{{\bf z}^H {\bf A} {\bf y}\}\right.\\
&\left.+\sum_{n=1}^N\lambda_n\left({\bf y}^He_{n+N(n-1)}e_{n+N(n-1)}^H{\bf y}-1\right)\right\},
\end{aligned}
\end{equation}
where ${\bf z}=\text{vec}({\bf G}\overline{\bf H}_1\overline{\bf H}_2)$.
Then, the optimal phase shift can be obtained as in Lemma 1.
\begin{lemma}
Given $\pmb\lambda=[\lambda_1, \lambda_2, \ldots, \lambda_N]$, the optimal phase shift of maximizing the Lagrange function, ${\cal L},$ is given by
\begin{equation}\label{s:16}
\begin{aligned}
&\phi_n^{(*)}=\\
&\frac{1}{\left(\underbrace{-{\bf A}\left({\bf A}+\sum_{n=1}^N{\lambda}_ne_{n+N(n-1)}e^H_{n+N(n-1)}\right)^{-1}   {\bf y}}_{\bar{\bf a}}\right)_{l(n)}},\\
& \forall n=1, 2, \ldots, N,
\end{aligned}
\end{equation}
where $l(n)=n+N(n-1)$ and $(\bar{\bf a})_{l(n)}$ represents the  $l(n)\text{th}$ element of vector $\bar{\bf a}$.
Then, the optimal dual variables $\pmb\lambda^{(*)}$ can be determined according to the constraints in (\ref{s:13}) via subgradient method \cite{Boyd2009Convex}.
\end{lemma}

{\textit{Proof:}}  The optimal $\phi_n^{(*)}$ in (\ref{s:16}) can be obtained by setting the first-order derivative of dual function to zero.

\subsection{The Optimal Transmit Power Allocation with Fixed $\pmb\Phi$}
In this subsection, we optimize the transmit power in (\ref{s:4}) given fixed $\pmb\Phi$.
Based on the ZF transmission design criterion, we have
${\bf W}=({\bf H}_1^H+{\bf H}_2^{H}{\pmb\Phi}{\bf G})^{\ddag}$.
Mathematically, the transmit power allocation can be formulated as the following optimization problem,
\begin{subequations}\label{s:17}
\begin{align}
&\max_{\bf P}~\sum_{k=1}^K\log_2\left(  1+p_k\sigma^{-2} \right)\\
\text{s.t.~}& ~~(\ref{add:1}\text{c})\text{~and~}(\ref{s:4}\text{b}).
\end{align}
\end{subequations}
Notably, for the nonlinear proportional rate constraints, it can be observed that for a given
${\pmb\xi}=\{\xi_1, \ldots, \xi_K\},$  a larger $\varphi$ will lead to a higher $R_k$, requiring more power consumption.
That is, the power matrix ${\bf P}$ is closely depended on $\varphi.$

Moreover, constraints (\ref{s:4}b) of the optimization problem (\ref{s:17}) are equivalent to the following equations
\begin{equation}\label{s:18}
\begin{aligned}
\trace&\left(\left({\bf H}_1^H+{\bf H}_2^H {\pmb\Phi}{\bf G}\right)^{\ddag}{\bf P}\left(  {\bf H}_1^H+{\bf H}_2^H{\pmb\Phi}{\bf G}\right)^{\ddag H} \right)\\
&\overset{\text{(a)}}{=}\trace\left({\bf W}{\bf Q}{\bf Q}^H{\bf W}^H\right)\\
&=\left\|{\bf W}{\bf Q} \right\|_F^2\\
&=\left\|\text{vec}\left(  {\bf W}{\bf Q}\right) \right\|^2\\
&=\left\| ({\bf I}\otimes {\bf W})\text{vec}({\bf Q}) \right\|^2\\
&=\text{vec}\left({\bf Q}\right)^H \left( {\bf I}\otimes {\bf W}\right)^H\left({\bf I}\otimes {\bf W}\right)\text{vec}({\bf Q})\leq P_{\max}.
\end{aligned}
\end{equation}
In (\ref{s:18}), we show (a) by the definitions ${\bf W}=\left({\bf H}_1^{H}+{\bf H}_2^H {\pmb\Phi}{\bf G}\right)^{\ddag}$ and ${\bf Q}{\bf Q}^H={{\bf P}}.$
Similar to the proof of Theorem 1, all the equalities in (\ref{s:18}) are suitable.
Then,  problem (\ref{s:17}) can be further transformed into
\begin{subequations}\label{s:19}
\begin{align}
&\max_{\bf P}~\sum\limits_{k=1}^K\log_2\left(1+p_k\sigma^{-2}  \right)\\
\text{s.t.}& \text{vec}\left({\bf Q}\right)^H \left( {\bf I}\otimes {\bf W}\right)^H\left({\bf I}\otimes {\bf W}\right)\text{vec}({\bf Q})\leq P_{\max},\\
& R_1:R_2:\cdots:R_K=\xi_1:\xi_2:\cdots:\xi_K,
\end{align}
\end{subequations}
which is an SE maximization problem with proportional rate constraints.
Let ${\bf B}=({\bf I}\otimes {\bf W})^H({\bf I}\otimes{\bf W})$ with ${\bf B}=\{b_{l(k),l(j)}\}_{ k,j\in{\cal K}}$, where $b_{l(k),l(j)}$ denote the $(l(k),l(j))\text{th}$ element of matrix ${\bf B}$
and $l(k)=k+K(k-1).$
The optimization problem in (\ref{s:19}) is equivalent to finding the maximum of the following Lagrangian function
\begin{equation}\label{ss:1}
\begin{aligned}
{\cal J}({\bf P}, &~{\pmb\mu})=\sum_{k=1}^K \log_2(1+p_k\sigma^{-2})
+\mu_1\bigg(P_{\max}\\
&-\sum_{k=1}^Kb_{l(k),l(k)}p_k
-2\sum_{k=1}^K\sum_{j>k}b_{l(k),l(j)}p_k^{\frac{1}{2}}p_j^{\frac{1}{2}} \bigg)\\
&+\sum_{k=2}^K\mu_k\bigg(\log_2(1+p_k\sigma^{-2})-\frac{\xi_k}{\xi_1}\log_2(  1+p_1\sigma^{-2})\bigg),
\end{aligned}
\end{equation}
where $\mu_k$ is the non-negative Lagrangian multiplier and ${\pmb\mu}$ is the collection of $\{\mu_1, \mu_2, \ldots, \mu_K\}.$
For user $k$, after differentiation it w.r.t. $p_k$ and setting the derivatives to zero,
we have
\setcounter{equation}{22}
\begin{equation}\label{ss:2}
\begin{aligned}
p_k^{(*)}=\left\{\begin{array}{lll}&\frac{1+\mu_k}{\ln2}\left(\mu_1b_{l(k),l(k)}+
\mu_1\left(\sum\limits_{j>k}^Kb_{l(k),l(j)}p_j^{\frac{1}{2}}\right.\right.\\
&\left.\left.+\sum\limits_{k>j}^Kb_{l(j),l(k)}p_j^{\frac{1}{2}}  \right)p_k^{-\frac{1}{2}} \right)^{-1}-\sigma^{2}, \forall k\geq 2;\\
&\frac{\frac{1}{\ln2}\left(1-\sum_{k=2}^K\frac{\mu_k\xi_k}{\xi_1}\right)}
{\mu_1\sum_{j= 1}^Kb_{1,l(j)}p_j^{\frac{1}{2}}p_1^{-\frac{1}{2}}}-\sigma^2, ~~~~~~~~~~~~~k=1.
\end{array}\right.
\end{aligned}
\end{equation}
By applying the subgradient method,  the dual variables $\mu_k, \forall k\in {\cal K},$ is optimally determined by
\setcounter{equation}{23}
\begin{equation}\label{sss:1}
\begin{aligned}
\mu_1^{(*)}=&\min\bigg\{ \mu_1\geq 0: \sum_{k=1}^Kb_{l(k),l(k)}p_k\\
&~~~~~~~+\sum_{k=1}^K\sum_{j>k}^K b_{l(k),l(j)}p_k^{\frac{1}{2}}p_j^{\frac{1}{2}}\leq P_{\max}\bigg\};\\
\mu_{k}^{(*)}=&\min\bigg\{\mu_k>0:  \log_2( 1+p_k\sigma^{-2})\\
&~~~~~~~-\frac{\xi_k}{\xi_1}\log_2(1+p_1\sigma^{-2})=0 \bigg\}, \forall k\geq 2.
\end{aligned}
\end{equation}
\begin{figure*}[!t]
\centering
\subfigure[The average SE vs  $P_{\max}.$]{
\label{fig:1}
\includegraphics[width=0.31\textwidth]{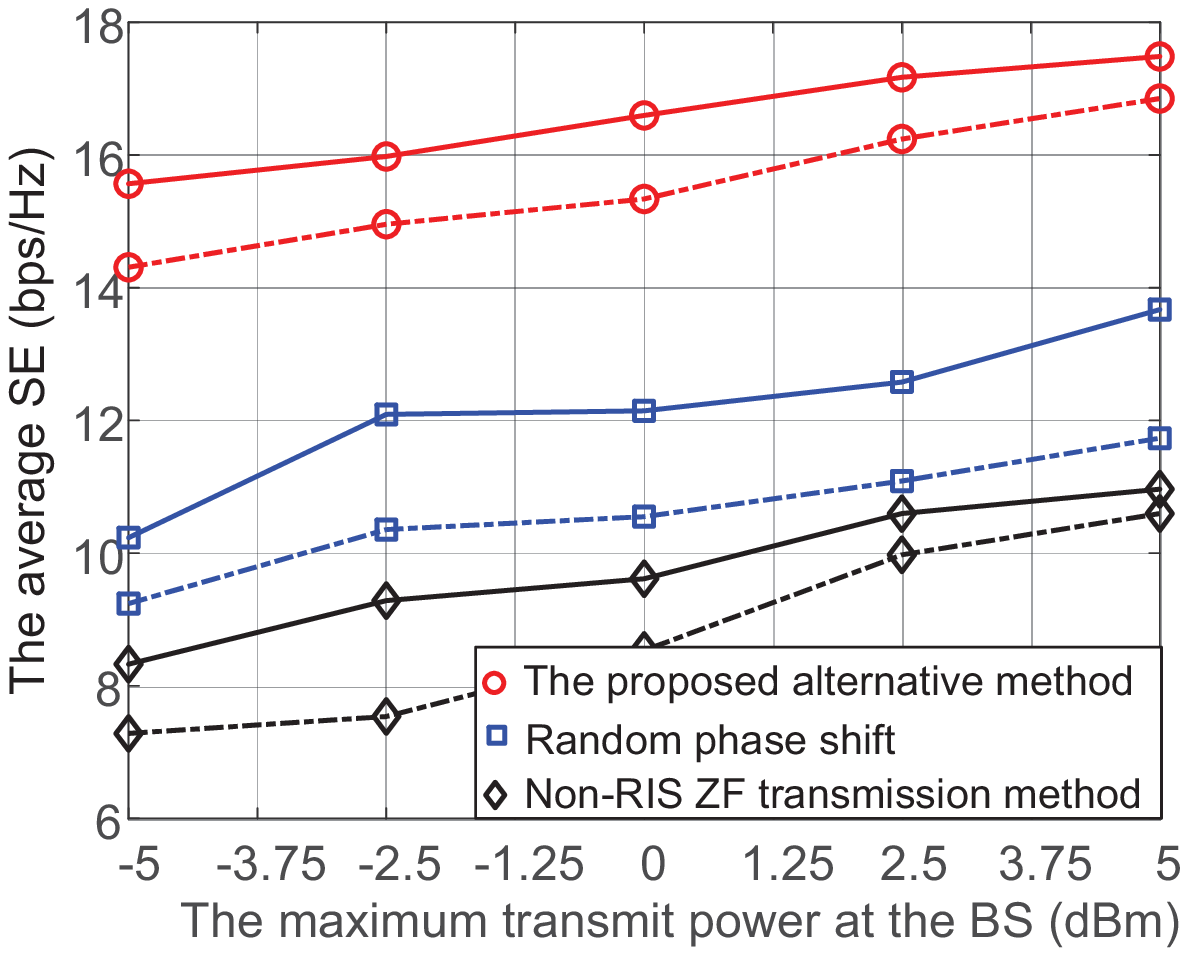}}
\subfigure[The average SE vs $N$.]{
\label{fig:2}
\includegraphics[width=0.32\textwidth]{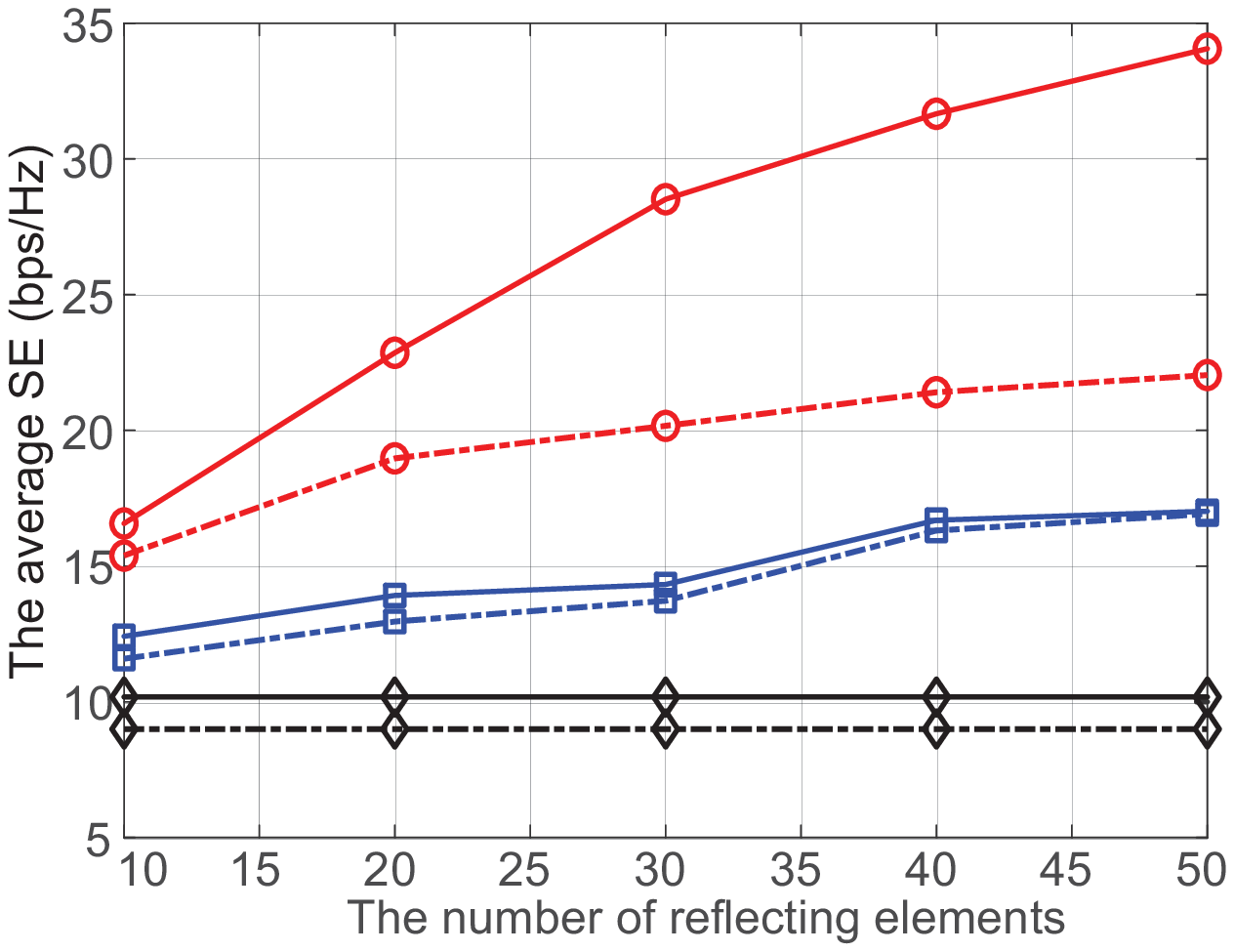}}
\subfigure[The average SE vs  $D$.]{
\label{fig:3}
\includegraphics[width=0.31\textwidth]{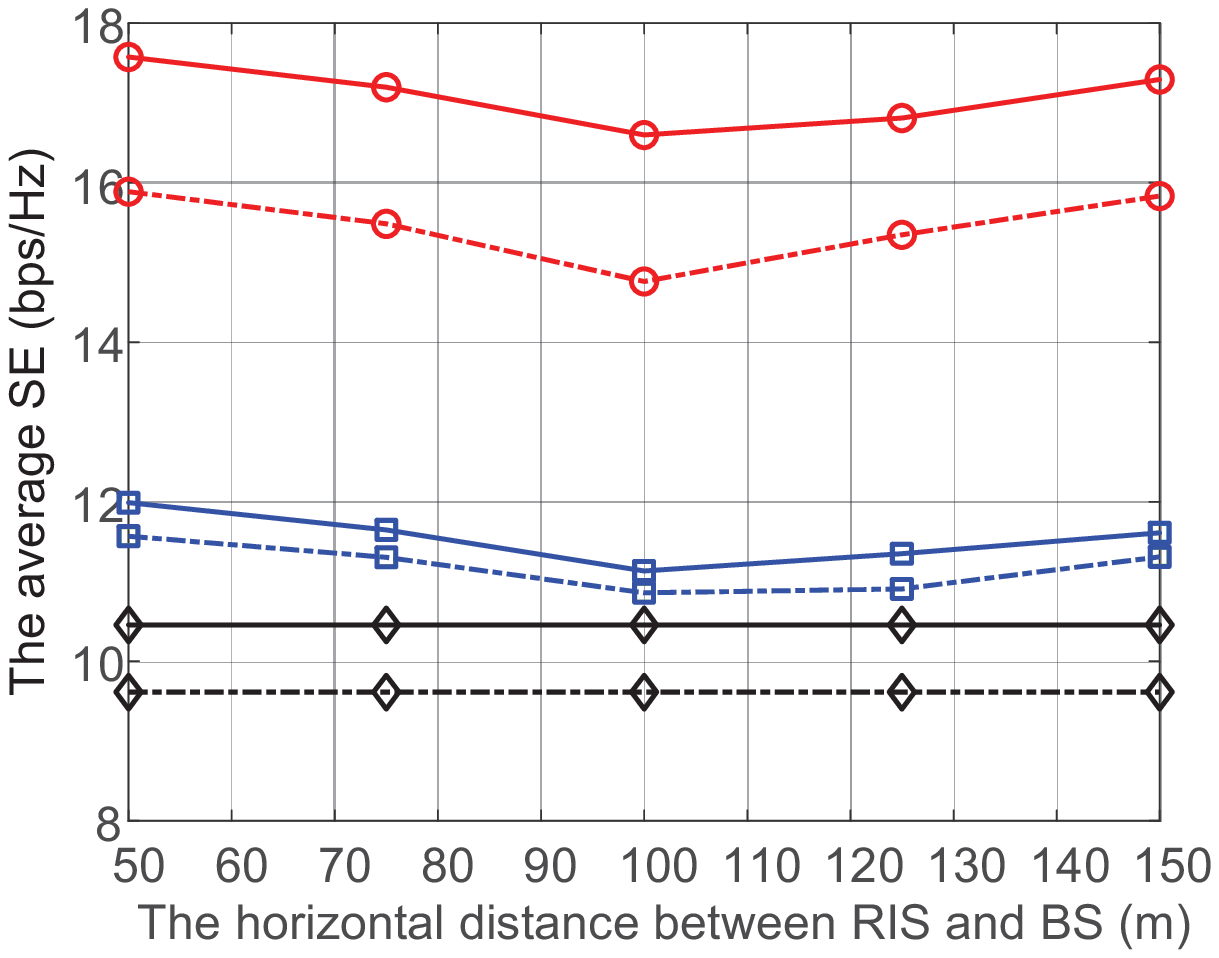}}
\caption{Performance comparisons between the proposed algorithm and two baselines.}
\label{fig:4}
\end{figure*}
\begin{lemma}
For a given $\varphi,$ the optimally allocated power associated with the $1\text{st}$ user is
\begin{equation}\label{ss:3}
\begin{aligned}
p_1(\varphi)=\frac{\frac{1}{\ln2}\left(1-\sum_{k=2}^K\frac{\mu_k\xi_k}{\xi_1}\right)}
{\mu_1\sum_{j= 1}^K b_{1,l(j)}\sqrt{\frac{2^{\varphi\xi_j}-1}
{2^{\varphi\xi_1}-1}}}-\sigma^2,
\end{aligned}
\end{equation}
and the optimal power allocation intended for  user $k, k\geq 2$ is
{\begin{equation}\label{ss:4}
\begin{aligned}
p_k&(\varphi)=\frac{1+\mu_k}{\ln2}\left(\mu_1b_{l(k),l(k)}+
\mu_1\left(\sum\limits_{j>k}^Kb_{l(k),l(j)}\right.\right.\\
&\left.\left.\sqrt{\frac{2^{\varphi\xi_j}-1}{2^{\varphi\xi_k}-1}}
+\sum\limits_{k>j}^Kb_{l(j),l(k)} \sqrt{\frac{2^{\varphi\xi_j}-1}
{2^{\varphi\xi_k}-1}}      \right)  \right)^{-1}-\sigma^2.
\end{aligned}
\end{equation}}
\end{lemma}

{\textit{Proof:} } Lemma 2 can be proved by substituting (\ref{eqq:1}) into (\ref{ss:2}).

Putting together the solutions for $\pmb\Phi$ and ${\bf P}$ presented respectively in Sections III-A and III-B, the complete alternatively algorithm for solving (\ref{add:1}) is summarized in Algorithm 1.
\begin{algorithm}

\caption{ The alternating algorithm for solving (\ref{add:1})}             

\label{alg:Framwork}                  

\begin{algorithmic}[1]                

 \STATE   The BS and RIS initialize ${\bf P}$ and ${\pmb\Phi}$, respectively.

 Set the predetermined proportional rate constraints $\pmb\xi.$

 {\bf Repeat}

\STATE {\bf Step 1: } Update the phase shift matrix at the RIS ${\pmb\Phi}=\text{diag}[\phi_1, \phi_2, \ldots, \phi_N]$ by (\ref{s:16}).

\STATE {\bf Step 2:} Update the transmit power at the BS by (\ref{ss:2}).

{\bf Until} The value of function (\ref{add:1}a) converges.

\end{algorithmic}

\end{algorithm}

\section{Simulation Results}



In line with \cite{Guo2019Weighted}, we consider an RIS-aided femtocell network with a two-dimension distribution model, in which the BS and RIS are respectively located at coordinates $(0, 0 )\text{~m}$ and $(D, 50 )\text{~m}.$
The BS equipped with $4$ antennas serves for $4$ single-antenna users which are randomly distributed within the special area, i.e., a circle centered at $(200, 0)\text{~m}$ with radius $10\text{~m}.$
Throughout the simulations, unless otherwise specified,  the transmission bandwidth is set to be $180\text{~kHz},$ and the noise power spectral density is $-174\text{~dBm/Hz}$ (see \cite{Gao2019Dynamic}, and references therein).
In the simulations, to include the effects of fading and shadowing, we use the path-loss model introduced in \cite{Gao2019Dynamic}, assuming perfect channel estimation at the BS and RIS as well as the use of ZF transmission technique.
In this section, the performance of the proposed alternative method in a multiuser MISO system is evaluated and compared with two baselines, denoted as random phase shift and Non-RIS ZF transmission method.
Specifically, 1) random phase shift: the proposed alternative method with optimal transmit power allocation, while the phase shift matrix of the RIS is not-optimized but randomly selected;
2) Non-RIS ZF transmission method: the conventional SE maximization with $N=0,$ then ZF transmission method is the optimal power allocation.
Figs. \ref{fig:1}-\ref{fig:3} depict the average SE for the three methods mentioned above in terms of $P_{\max}$, $N,$ and $D$, respectively, averaged over $500$ independent channel realizations per marker.
And the rate fairness constraints among the $4$ users are respectively set to be $\xi_1:\xi_2:\xi_3:\xi_4=1:1:1:1$ (solid lines) and  $\xi_1:\xi_2:\xi_3:\xi_4=1:2:3:4$ (dotted lines).

Fig. \ref{fig:1} plots the average  SE achieved by all above methods versus the maximum transmit power of the BS $P_{\max}$.
In Fig. \ref{fig:1},  $N=10, D=100,$ the reflecting efficiency $\eta=0.8$,
From Fig. \ref{fig:1}, we can see that the proposed alternative method always achieves SE performance gain compared with random phase shift and Non-RIS ZF transmission methods.
As expected, the average SE of all the mentioned methods increases for a given value of ${\pmb\xi}$, as the maximum transmit power, $P_{\max},$ increases, and the proposed alternative method always outperforms the other two methods.
In addition to our observations in Fig. \ref{fig:1} with respect to the $P_{\max}$, the average SE achieved by all the mentioned methods various as the rate constraint changes.
Notably, by setting $\xi_1:\xi_2:\cdots\xi_K=1:1:\cdots:1,$ the objective of the optimization in (\ref{add:1}) is identical to the problem of maximizing the minimum user's achievable rate, since the worst user's achievable rate is maximized when all users have the same rate and SE is maximized.
Thus, the maximizing minimum  user's rate problem is a special case of the framework presented in this paper.

Fig. \ref{fig:2} shows the influence of the number of reflecting elements of RIS on SE with $M=4, K=4, $ and $P_{\max}=0\text{~dBm}$, varying $N $ from $10 $ to $50.$
From the results, we observe that for all the mentioned methods with the aid of RIS, the average SE increases, when the number of reflecting elements $N$ increases.
This is mainly because that the sum power of signals reflected by the RIS becomes stronger.
However, the increase of SE for the proposed alternative method due to increasing the number of reflecting elements $N$ is more significant, than that for random phase shift method.

Fig. \ref{fig:3} represents the average SE performance of the above three methods against the deployment of RIS, in this paper, the influence of the horizontal coordinate of the RIS, $D$, is focused on.
In Fig. \ref{fig:3}, we set $M=4, K=4, N=10,$ and the maximum transmit power of the BS $P_{\max}=0 \text{~dBm}.$
It is clear from Fig. \ref{fig:3} that an appropriate choice of the horizontal coordinate $D$ of the RIS can lead to severe increase of the average SE, and the performance gain is highly sensitive to the position (i.e., $D$) of the RIS.
Specifically, the average  SE of the RIS-aided system increases when the RIS is deployed closer to the BS or the cluster of users, and deploying RIS at the center place ($D=100$) is the worst case.
This in essence attributes to the double-fading path loss model of the RIS-aided system.

\section{Conclusions}
In this paper, we studied the SE performance in RIS-aided multiuser MISO systems with ZF transmission method employed at the BS.
In order to explore the SE performance with user fairness for such a system, we formulated an optimization problem to maximize the SE by jointly optimizing the transmit power allocation at the BS and the phase shift at the RIS under nonlinear proportional rate constraints.
To attain the optimal solution of the problem, we proposed an alternatively iterative algorithm and its convergence can be verified by convex optimization theory easily.
Simulation results demonstrated that the proposed alternative method achieves significant performance gain compared with various baselines, i.e., the random phase shift method and the conversional ZF transmission method.

\section*{Appendix A\\ Proof of Theorem 1}

By expressing the power allocation matrix ${\bf P}(\varphi)={\bf Q}(\varphi){\bf Q}^H(\varphi)$ with ${\bf Q}\triangleq \sqrt{\bf P},$ we observe that the objective function in (\ref{s:10}) can be rewritten as
\begin{equation}\label{s:12}
\begin{aligned}
\hspace{-1em}{\cal F}(\pmb\Phi)\hspace{-0.2em}=&\trace\left(({\bf H}_1^H+{\bf H}_2^{H}{\pmb\Phi}{\bf G})^{\ddag}{\bf P}(\varphi)
({\bf H}_1^H+{\bf H}_2^H{\pmb\Phi}{\bf G})^{\ddag H} \right)\\
\overset{\text{(a)}}{=}&\left\|\left({\bf Q}^{-1}(\varphi)\left({\bf H}_1^H+{\bf H}_2^{H}{\pmb\Phi}{\bf G}\right)\right)^{\ddag} \right\|_F^2\\
\overset{\text{(b)}}{=}&\left\|{\bf G}^{\ddag}\left({\bf G}\overline{\bf H}_1^{\ddag}\overline{\bf H}_2+
{\pmb\Phi}^{-1}\right)\overline{\bf H}_2^{\ddag}\right\|^2_F\\
\overset{\text{(c)}}{=}&\left\| \text{vec}\left({\bf G}^{\ddag}\left({\bf G}\overline{\bf H}_1^{\ddag}
\overline{\bf H}_2+{\pmb\Phi}^{-1}\right) \overline{\bf H}_2^{\ddag}\right) \right\|^2\\
\overset{\text{(d)}}{=}&\left\|\left(\overline{\bf H}_2^{\ddag H}\otimes{\bf G}^{\ddag}\right)
\text{vec}\left({\bf G}\overline{\bf H}_1^{\ddag}\overline{\bf H}_2 +{\pmb\Phi}^{-1} \right)
\right\|^2\\
\overset{\text{(e)}}{=}&\text{vec}\left( \pmb\Phi^{-1}\right)^H
\left(\overline{\bf H}_2^{\ddag H}\otimes{\bf G}^{\ddag}\right)^H\left(\overline{\bf H}_2^{\ddag H}\otimes{\bf G}^{\ddag}\right)\\
&\text{vec}(\pmb\Phi^{-1})
+2\text{Re}\left\{\text{vec}\left({\bf G}\overline{\bf H}_1^{\ddag}\overline{\bf H}_2\right)^H\right.\\
&\left.\left(\overline{\bf H}_2^{\ddag H}\otimes {\bf G}^{\ddag}\right)^H
\left(\overline{\bf H}_2^{\ddag H}\otimes{\bf G}^{\ddag }\right)\text{vec}\left(\pmb\Phi^{-1}\right)\right\}\\
&+\text{vec}({\bf G}\overline{\bf H}_1^{\ddag}\overline{\bf H}_2)^H
\left(\overline{\bf H}_2^{\ddag H}\otimes{\bf G}^{\ddag }\right)^H\\
&~~~~\underbrace{\left(\overline{\bf H}_2^{\ddag H}\otimes {\bf G}^{\ddag }\right) \text{vec}({\bf G}\overline{\bf H}_1^{\ddag}\overline{\bf H}_2)}_{{\cal C}_0}.
\end{aligned}
\end{equation}
In the expression, whereas step (a) follows the properties of Frobenius norm and Pseudo-inverse law of a matrix product, the definitions $\overline{\bf H}_1={\bf Q}(\varphi){\bf H}_1^H$ and $ \overline{\bf H}_2={\bf Q}(\varphi){\bf H}_2^{H}$ used in step (b),
(c) follows from the vectorization operator, (d) follows from the law of $\text{vec}({\bf ABC})=({\bf A}\otimes {\bf C}^H)\text{vec}({\bf B}^H),$ and the law of $\text{vec}({\bf A}+{\bf B})=\text{vec}({\bf A})+\text{vec}({\bf B})$ and the symmetry of matrix $(\overline{\bf H}_2^{\ddag H}\otimes{\bf G}^{\ddag })^H(\overline{\bf H}_2^{\ddag H}\otimes {\bf G}^{\ddag}),$ respectively. Moreover, the final term of (\ref{s:12}), i.e., the term ${\cal C}_0$ is constant, thus, the optimization problem (\ref{s:11}) is equivalent to (\ref{s:12}).


\end{document}